\documentstyle[12pt]{article}

\global\arraycolsep=1pt
\newcommand{\beq}{\begin{equation}}
\newcommand{\eeq}{\end{equation}}
\newcommand{\beqa}{\begin{eqnarray}}
\newcommand{\eeqa}{\end{eqnarray}}
\newcommand{\CR}{\nonumber \\}

\renewcommand{\theequation}{\thesection.\arabic{equation}}
\renewcommand{\thefootnote}{\fnsymbol{footnote}}

\begin{document}

\begin{titlepage}
\begin{flushright}
{November, 2001} \\
{\tt hep-th/0111198}
\end{flushright}
\vspace{0.5cm}
\begin{center}
{\Large \bf
On $Spin(7)$ holonomy metric based on $SU(3)/U(1)$ : II
}%
\lineskip .75em
\vskip1.0cm
{\large Hiroaki Kanno\footnote{e-mail: kanno@math.nagoya-u.ac.jp}}
\vskip 1.0em
{\large\it Graduate School of Mathematics \\
Nagoya University, Nagoya, 464-8602, Japan}
\vskip 0.8cm
{\large Yukinori Yasui\footnote{e-mail: yasui@sci.osaka-cu.ac.jp}}
\vskip 1.0em
{\large\it Department of Physics, Osaka City University \\
Sumiyoshi-ku, Osaka, 558-8585, Japan}
\end{center}
\vskip0.5cm

\begin{abstract}

We continue the investigation of $Spin(7)$ holonomy metric of
cohomogeneity one with the principal orbit $SU(3)/U(1)$.
A special choice of $U(1)$ embedding in $SU(3)$ allows more
general metric ansatz with five metric functions. There are two 
possible singular orbits in the first order system of $Spin(7)$ 
instanton equation. One is the flag manifold $SU(3)/T^2$ also 
known as the twister space of ${\bf CP}(2)$ and the other is 
${\bf CP}(2)$ itself. Imposing a set of algebraic constraints,
we find a two-parameter family of exact solutions which have 
$SU(4)$ holonomy and are asymptotically conical. There are 
two types of asymptotically locally conical (ALC) metrics in our model,
which are distingushed by the choice of $S^1$ circle whose radius
stabilizes at infinity. We show that this choice of $M$ theory circle
selects one of possible singular orbits mentioned above.
Numerical analyses of solutions near the
singular orbit and in the asymptotic region support
the existence of two families of ALC $Spin(7)$ metrics:
one family consists of deformations of the Calabi hyperK\"ahler metric, 
the other is a new family of metrics on
a line bundle over the twister space of ${\bf CP}(2)$.


\end{abstract}
\end{titlepage}

\renewcommand{\thefootnote}{\arabic{footnote}}
\setcounter{footnote}{0}

\section{Introduction}
\setcounter{equation}{0}

By string dualities intriguing dynamics in supersymmetric
compactification of superstrings and $M$ theory are often
associated with singularities in manifolds of special
holonomy which appear at finite distance in the moduli space.
If the singularity is isolated and conical, we may expect
that the details of the metric far from the singularity are
irrelevant and as an approximation of the singular geometry
take a simple Ricci-flat cone metric over an $n$-dimensional
Einstein manifold $M$;
\beq
ds^2 = dr^2 + r^2 d\Omega^2~,  \qquad (0 \leq r < +\infty)
\eeq
where the Einstein metric $d\Omega^2$ satisfies $R_{ab}=(n-1)g_{ab}$.
For supersymmetry a parallel spinor should exist on the cone
and it comes from a Killing spinor on $M$ \cite{Bar}, \cite{AFHS}, \cite{MP}.
Unless the Einstein manifold is the $n$-dimensional
sphere $S^n = SO(n+1)/SO(n)$, there is a conical singularity at
$r=0$. When the manifold $M$ is a homogeneous space $G/K$,
a resolution of the singularity may be provided by
the following metric of cohomogeneity one \cite{DW}, \cite{Wang}, \cite{EW};
\beq
d\widetilde{s}^2 = dt^2 + g_{G/K} (t)~,  \qquad  ( t_0 \leq t < +\infty)
\eeq
where  $g_{G/K}(t)$ is a $t$-dependent homogeneous metric
on the principal orbit $G/K$. This sort of resolution of
an isolated conical singularity has been employed recently
in the discussion of IR strong coupling dynamics of supersymmetric
gauge theories based on gauge theory/gravity
correspondence in large $N$ limit \cite{KS}, \cite{MN}, \cite{Vafa},
\cite{Ach}, \cite{AMV}, \cite{PZT}, \cite{EN}, \cite{AW}.

The requirement of special holonomy, which can be expressed
as linear constraints on the spin connection $\omega_{ab}$, gives
a first order system of flow equations for one-parameter family
of homogeneous metrics $g_{G/K}(t)$.
The boundary condition should be specified in solving the flow
equation. At the boundary $t=t_0$ there appears a singular
orbit $G/H$ with $K \subset H \subset G$. This singular orbit has a finite
volume and the original conical singularity is developed when the volume
of $G/H$ tends to vanishing. The coset $H/K$ has to be
a sphere $S^k$ for the principal orbit $G/K$ to degenerate
smoothly to the singular orbit \cite{Mos}. When there are several choices of
the subgroup $H$ such that $H/K \simeq S^k$,
there may be more than one way of
resolving the conical singularity. A famous example is given by
the conifold that is a cone over the five dimensional coset
space $T^{1,1} = SU(2) \times SU(2) / U(1)$.
There are three possible singular orbits \cite{CGLP2};
\begin{enumerate}
\item
$H= U(2)~, \quad G/H \simeq S^2~, \quad  H/K \simeq S^3~,$
\item
$H= SU(2)~, \quad G/H \simeq S^3~, \quad  H/K \simeq S^2~,$
\item
$H= U(1) \times U(1)~, \quad G/H \simeq S^2 \times S^2~, \quad
H/K \simeq S^1~.$
\end{enumerate}
In this paper we will see a similar example of this kind, when the principal
orbit is the seven dimensional coset space $N (1,1) = SU(3)/ U(1)$.

The other side of the boundary is specified by the asymptotic behavior
of the solution.
A standard behavior is that the homogeneous metric $g_{G/K}(t)$ asymptotically
approaches to the original Einstein metric $d\Omega^2$.
Such metrics are called asymptotically conical (AC). From the viewpoint of
compactifications of $M$ theory we are also interested in
the asymptotic behavior called asymptotically locally conical (ALC) 
\cite{CGLP3},
where there is a circle whose radius remains finite at infinity .
In \cite{Gom}, by considering the geometry of ALE fibration over
a supersymmetric cycle, it has been argued that an $M$ theoretic lift
of a type IIA geometry with $D6$ branes wrapping on the SUSY cycle
is given by purely gravitational configuration. (See also \cite{GS}
for a relation of the $M$ theoretic lift to $Spin_c$ structure.)
Such $D6$ brane configurations have been discussed from the
dual picture of eight dimensional supergravity in \cite{Her}, \cite{GM2}.
Since the $M$ theory circle which is related to
the string coupling of IIA theory should remain finite asymptotically,
the corresponding metric is expected to be ALC. In fact
when the SUSY cycle is $S^4$ in $Spin(7)$ manifold and $S^3$ in $G_2$ manifold,
such ALC metrics were constructed in \cite{CGLP3} and \cite{BGGG}, 
respectively.
More recently a similar ALC metric has been found for a SUSY cycle
${\bf CP} (2)$ in \cite{KY}, \cite{CGLP5}, \cite{GS}.
Even if we assume that the metric is ALC, the choice of $M$ theory circle
in the principal orbit may not be unique, when there are more than one 
irreducible
modules of dimension one in the isotropy representation on the tangent space
of the principal orbit.
Due to the special choice of $U(1)$ subgroup to be introduced shortly,
the isotropy representation of the coset space $N(1,1) = SU(3)/U(1)$ has
three one dimensional irreducible components.
Recently $M$ theory on ALC $Spin(7)$ manifolds has been discussed in 
\cite{GS}, \cite{CKL}.

In this article taking the homogeneous space $SU(3)/U(1)$ (also known as
the Aloff-Wallach space), we investigate aspects of the special holonomy
metrics of cohomogeneity one.
In our previous work \cite{KY} we left a choice of $U(1)$ subgroup in
$SU(3)$ free
so that the triality $W(SU(3))$(=the Weyl group) symmetry was manifest.
In the following we will fix the embedding so that the $U(1)$ subgroup is
${\rm diag} (e^{i\theta}, e^{i\theta}, e^{-2i\theta})$. In this
case we can make a more general metric ansatz with five functions,
while in general the number is four. In section two we derive
a first order system for $Spin(7)$ holonomy and classify a possible singular
orbit appearing at the boundary.  In our metric ansatz there is a natural
candidate for a K\"ahler two form. The closedness (or the integrability) of
the candidate two form gives a set of algebraic constraints
that allows us to solve the flow equation exactly.
In section three we present a two-parameter family of exact solutions
which is asymptotically conical.
They are $SU(4)$ holonomy metrics
on the line bundle over the flag manifold $SU(3)/T^2$, which is a
two-sphere bundle
over ${\bf CP}(2)$. When one of the parameters vanishes,
then the $S^2$ fiber collapses
and the metric reduces to the Calabi hyperK\"ahler metric on $T^* {\bf CP}(2)$.
An analysis of ALC solutions is carried out in section four.
Due to the generalized metric ansatz with five metric
functions there are two choices of a circle whose radius remains finite 
asymptotically.
We find that if we assume the metric is non-singular,
the ALC asymptotic behavior requires a reduction of
the number of metric functions from five to four, but the way of reduction
depends on the choice of $S^1$ factor in the principal orbit.
 From the perturbative analysis around the singular orbit
we see one of possible singular orbits is selected by each reduction
and thus there are two types of ALC $Spin(7)$ metrics.
The topology of the singular orbit and the choice of
$M$ theory circle cannot be independent and the singular orbit is
either ${\bf CP}(2)$ or $Flag_6 = SU(3)/T^2$
depending on the chioce of asymptotic $M$ theory circle.

Since we could not find explicit solutions in general, we have numerically
worked out perturbative series expansions both
around the singular orbit and in the asymptotic region.
In section five, based on this numerical analysis
we propose the \lq\lq moduli\rq\rq\ space
of $Spin(7)$ metrics of cohomogeneity one with the principal orbit
$SU(3)/U(1)$ for the special choice of $U(1)$ subgroup.
Especially we observe
that two types of ALC metrics in section four are in fact interpolated
by the exactly known Ricci-flat K\"ahler metrics obtained in section three.
The existence of ALC metrics whose singular orbit is $Flag_6$ is
shown only numerically. But their qualitative behavior is much like
the Atiyah-Hitchin metric in four dimensions. Hence we believe that
this is an analogue of $Spin(7)$ metric called ${\bf C}_8$ in \cite{CGLP5},
\cite{CGLP6}, whose singular orbit is ${\bf CP}(3)$, the twister space
of $S^4$.

\section{Instanton equation with five metric functions}
\setcounter{equation}{0}

The maximal torus $T^2$ of the Lie group $SU(3)$ is two dimensional
and its $U(1)$ subgroup is specified by
integers $\overrightarrow n = (n_1, n_2, n_3)$ with $n_1 + n_2 + n_3=0$.
Without loss of generality we can assume that there is no
common divisor of $n_i$. Explicitly the $U(1)$ subgroup is given by
${\rm diag} (e^{in_1\theta}, e^{in_2\theta}, e^{in_3\theta})$.
In previous paper on $Spin(7)$ metric of cohomogeneity one
with the principal orbit $SU(3)/U(1)$, we took the following
metric ansatz \cite{KY};
\beq
g = dt^2 + a(t)^2 (\sigma_1^2 + \sigma_2^2) + b(t)^2
              (\Sigma_1^2 + \Sigma_2^2) + c(t)^2 (\tau_1^2 + \tau_2^2)
+ f(t)^2 T_A^2~,
\eeq
which is consistent with any choice of the embedding parameters
$\overrightarrow n$ and consequently gives a formulation
which has manifest symmetry under $\Sigma_3 = W(SU(3))$;
the Weyl group of $SU(3)$. Our convention of $SU(3)$ left invariant
one forms is summarized in Appendix A. The components of
invariant one form for the maximal torus $T^2$
are denoted as $T_A$ and $T_B$. The corresponding generators are given by
\beq
E_A = -\frac{1}{\Delta} \left(
\begin{array}{ccc}
\alpha_B & 0 & 0 \\
0 & \beta_B & 0 \\
0 & 0 & \gamma_B
\end{array}
\right),  \qquad
E_B=\frac{1}{\Delta} \left(
\begin{array}{ccc}
\alpha_A & 0 & 0 \\
0 & \beta_A & 0 \\
0 & 0 & \gamma_A
\end{array}
\right),  \label{base}
\eeq
with $\alpha_A + \beta_A + \gamma_A = \alpha_B + \beta_B + \gamma_B =0$
and $\Delta = \beta_A \alpha_B - \alpha_A \beta_B$.
The generator of the $U(1)$ subgroup is $E_B$.
Note that the bases of the Lie algebra $su(3)$ and the components
of the left invariant one form are in dual relation and hence
the role of parameters $\alpha_A, \beta_A, \gamma_A$ and
$\alpha_B, \beta_B, \gamma_B$ is exchanged.

When the $U(1)$ subgroup generated by $E_B$ decouples from
one of $\sigma$, $\Sigma$ and $\tau$, more general metric
ansatz is allowed since in this case the isotropy representation
of $SU(3)/U(1)$ becomes
\beq
su(3)/u(1)={\bf p}_1 \oplus {\bf p}_2 \oplus {\bf p}_3 \oplus
\widetilde{{\bf p}_3} \oplus
{\bf p}_4,
\eeq
where $\mbox{dim}~{\bf p}_1=
\mbox{dim}~{\bf p}_2 =2$
and $\mbox{dim}~{\bf p}_3 =
\mbox{dim}~\widetilde{{\bf p}_3} =
\mbox{dim}~{\bf p}_4=1$.
Then the metric ansatz becomes
\beq
g = dt^2 + a(t)^2 (\sigma_1^2 + \sigma_2^2) + b(t)^2
              (\Sigma_1^2 + \Sigma_2^2) + c_1(t)^2 \tau_1^2 +  c_2(t)^2 \tau_2^2
+ f(t)^2 T_A^2~, \label{ansatz}
\eeq
where we assume that $\tau_1$ and $\tau_2$ are singlets.
The reduction of the holonomy group from $SO(8)$ to $Spin(7)$ is represented by
the octonionic instanton equation \cite{CDFN}, \cite{AL}, \cite{BFK},
(See also Appendix B).
We can see the octonionic instanton equation derived
from the above ansatz does not have $T_B$ component,
if and only if $d\tau_i$ has no $T_B$ component.
We have $\nu_B =0$ and hence $\alpha_A=\beta_A$ (see Appendix A).
Then the generator of the $U(1)$ subgroup is
$E_B ={\rm diag} (1, 1, -2)$ and
the charge vector $\overrightarrow n$ in the
Maurer-Cartan equation of $dT_A$
is fixed to be $\overrightarrow n = (1,1,-2)$.
Note that this is the case where the action of the Weyl group degenerates.
We obtain the following system of first order differential equations
as the octonionic instanton equation on the spin connection $\omega_{ab}$
derived from the metric ansatz (\ref{ansatz});
\beqa
\frac{\dot a}{~a~} &=& \frac{b^2 + c_1^2 - a^2}{2abc_1}
+ \frac{b^2 + c_2^2 - a^2}{2abc_2} - \frac{f}{a^2}~, \CR
\frac{\dot b}{~b~} &=& \frac{c_1^2 + a^2 - b^2}{2abc_1}
+ \frac{c_2^2 + a^2 - b^2}{2abc_2}- \frac{f}{b^2}~, \CR
\frac{\dot c_1}{~c_1~} &=& \frac{a^2 + b^2 - c_1^2}{abc_1} +
\frac{2f}{c_1 c_2} + \frac{c_2^2 -c_1^2}{2c_1 c_2 f}~, \label{five} \\
\frac{\dot c_2}{~c_2~} &=& \frac{a^2 + b^2 - c_2^2}{abc_2} +
\frac{2f}{c_1 c_2} + \frac{c_1^2 -c_2^2}{2c_1 c_2 f}~, \CR
\frac{\dot f}{~f~} &=& \frac{f}{a^2}
+ \frac{f}{b^2} - \frac{2f}{c_1 c_2} + \frac{(c_1- c_2)^2}{2c_1 c_2 f}~.
\nonumber
\eeqa
This first order system has a discrete ${\bf Z}_2 \times {\bf Z}_2$ symmetry
generated by $(a,b,c_1,c_2) \to (\pm a, \mp b, -c_1, -c_2)$ and two exchange
symmetries $a \leftrightarrow b$ and $c_1 \leftrightarrow c_2$.
We note that though any independent sign flip of metric functions that
is not necessarily included above has no effect on the metric itself
or at the level of Ricci-flatness, but do {\it not} keep
the instanton equation invariant.
The first order system (\ref{five}) is an integral of
the second order Einstein equation
and the change in the instanton equation
means the different ways of integration.

Let us classify possibilities of the singular orbit compatible with
the evolution equation (\ref{five}). Group theoretically the singular
orbit is in one to one correspondence to a subgroup $H$ which
satisfy $U(1) \subset H \subset SU(3)$ and $H/U(1)$ should be
a sphere which is collapsing at the singular orbit.
Thus we find three possibilities;
\begin{enumerate}
\item
$H = U(1)\times U(1)$ ; In this case $H/U(1) \simeq S^1$ is collapsing and
the singular orbit is the twister space of ${\bf CP}(2)$; $SU(3)/H \simeq
Flag_6$, which is topologically a two sphere bundle over ${\bf CP}(2)$.
\item
$H = SU(2)$ ; In this case $H/U(1) \simeq S^2$ is collapsing and
the singular orbit is $SU(3)/H \simeq S^5$, which is the
Hopf bundle over ${\bf CP}(2)$.
\item
$H= S(U(2) \times U(1))$ ; In this case $H/U(1) \simeq S^3$ is collapsing and
the singular orbit is $SU(3)/H \simeq {\bf CP}(2)$ itself.
\end{enumerate}
We assume that the singular orbit is at $t=0$ and
make the following series expansion for small $t$;
\beqa
a(t) &=& p + \sum_{k \geq 1} a_k t^k~, \qquad
b(t) = q + \sum_{k \geq 1} b_k t^k~, \CR
c_1(t) &=& m + \sum_{k \geq 1} c_{1k} t^k~, \qquad
c_2(t) = n + \sum_{k \geq 1} c_{2k} t^k~, \CR
f(t) &=& r + \sum_{k \geq 1} f_k t^k~.
\eeqa
The parameters $p,q,m,n,r$ are regarded as the \lq\lq initial
conditions\rq\rq\ at
the singular orbit.
Substituting the series expansion to (\ref{five}) and looking at
the leading order, we obtain
\beqa
a_1 &=& \frac {1}{2}
\left( \frac{q}{m} + \frac{m}{q} - \frac{p^2}{mq} +
\frac{q}{n} + \frac{n}{q} - \frac{p^2}{nq} -\frac{2r}{p} \right)~, \CR
b_1 &=& \frac{1}{2}
\left(  \frac{p}{m} + \frac{m}{p} - \frac{q^2}{mp} +
\frac{p}{n} + \frac{n}{p} - \frac{q^2}{np} -\frac{2r}{q} \right)~, \CR
c_{11} &=& \frac{p}{q} + \frac{q}{p} - \frac{m^2}{pq} +\frac{2r}{n}
+\frac{n}{2r} - \frac{m^2}{2nr}~, \\
c_{21} &=& \frac{p}{q} + \frac{q}{p} - \frac{n^2}{pq} +\frac{2r}{m}
+\frac{m}{2r} - \frac{n^2}{2mr}~, \CR
f_1 &=& \frac{r^2}{p^2} + \frac{r^2}{q^2} -\frac{2r^2}{mn}
+ \frac{n}{2m} + \frac{m}{2n} -1~. \nonumber
\eeqa
Now the above three possibilities of the singular orbit correspond
respectively to the following initial conditions;
\begin{enumerate}
\item
$S^1$ is collapsing; $r=0$~,
\item
$S^2$ is collapsing; $p=0,~{\rm or}~~q=0$~,
\item
$S^3$ is collapsing; $p=r=0,~{\rm or}~~q=r=0$~.
\end{enumerate}
Firstly in case 2 there is no
regular solution at the singular orbit, since
there is no way to make $f_1$ regular\footnote{However,
another special choice of $U(1)$ embedding seems to allow
$S^5$ as a singular orbit \cite{CGLP5}. It might be very interesting
to see why it is the case,
since there is no odd dimensional SUSY cycle
in eight dimensions.}.
On the other hand in case 1
we see that the regularity of $c_{11}$ and $c_{21}$ requires
$m^2=n^2$. But $m=n$ implies $f_1=0$, which means that
the $S^1$ is collapsing \lq\lq too\rq\rq\ fast near $t=0$.
Thus only the case $m=-n$ can give non-singular
solutions. This also means that this type of singular orbit is
not allowed in generic cases where we have $c_1=c_2$.
In this case $p$ and $q$ are free parameters and $f_1=-2$.
Finally in case 3 there are non-singular
solutions if $m^2=n^2=q^2$ or $m^2=n^2=p^2$. Thus near $t=0$ we have
two types of
boundary conditions which correspond to case 1 and case 3, respectively~;
\beqa
g & \longrightarrow & dt^2+ 4t^2 T_{A}^2 + p^2(\sigma_{1}^2+\sigma_{2}^2)
+q^2(\Sigma_{1}^2+\Sigma_{2}^2)+ m^2(\tau_{1}^2+\tau_{2}^2)~, \label{ab} \\
g & \longrightarrow & dt^2+ t^2(T_{A}^2 + \sigma_{1}^2 + \sigma_{2}^2)
+m^2(\Sigma_{1}^2 + \Sigma_{2}^2 + \tau_{1}^2 + \tau_{2}^2)~. \label{bb}
\eeqa
Note that the terms with $\sigma_i^2$, $\Sigma_i^2$ and $\tau_i^2$ in 
(\ref{ab})
describe the singular orbit $Flag_6$ squashed by the parameters $p,q$ and $m$,
while the term with $\Sigma_i^2+\tau_i^2$ in (\ref{bb}) represents
the singular orbit ${\bf CP}(2)$ with the Fubini-Study metric.
In the following we call the first case
A-type boundary and the second case B-type boundary.
If we regard the homogeneous space $Flag_6$ as an $S^2$-bundle over
${\bf CP}(2)$, then B-type boundary may be reduced from A-type one
by making the fibre $S^2$ collapse. The higher order terms of the series
expansion are summarized in Appendix C.

\section{Explicit AC solutions of $SU(4)$ holonomy}
\setcounter{equation}{0}

In terms of the vielbeins (the orthonormal frames) of our metric ansatz,
we can write down the following non-degenerate two form
\beq
\omega=fdt \wedge T_{A} - c_1 c_2 \tau_1 \wedge \tau_2
-a^2 \sigma_1 \wedge \sigma_2 -b^2 \Sigma_1 \wedge \Sigma_2~,
\eeq
which is a natural candidate for a K\"ahler form.
Using the first order equation for functions  $a,b$ and $c_i$,
we see that the condition $d\omega =0$ is equivalent to
the constraints
\beq
a^2 + b^2 + c_1 c_2 = 0~, \quad c_1+c_2=0~.
\eeq
They are compatible with the first order system (\ref{five}) and
we obtain the following reduction with $c:= c_1 = - c_2$~;
\beqa
\dot a &=& -\frac{f}{a}~, \quad \dot b =-\frac{f}{b}~, \quad
\dot c =-\frac{2f}{c}~, \label{first} \\
\dot f &=& \frac{f^2}{a^2}
+ \frac{f^2}{b^2} + \frac{2f^2}{c^2} - 2~.
\label{second}
\eeqa
We can solve this reduced first order system exactly.
The first two equations of (\ref{first}) implies
\beq
b^2 - a^2 = \ell^2~,
\eeq
where $\ell^2$ is an integration constant. Due to the exchange
symmetry of $a$ and $b$, we may assume $b^2 - a^2 \geq 0$.
Furthermore by a change of variables $dt = c/(2f) dr$ we can integrate
$a,b,c$ to obtain
\beq
a^2 = \frac{1}{2} (r^2 - \ell^2)~, \quad  b^2 = \frac{1}{2} ( r^2 + \ell^2 )~,
\quad c^2 = r^2 ~,
\eeq
where we have fixed an integration constant by requiring $c^2=r^2$.
Substituting the above solution into the equation (\ref{second})
we have
\beq
\frac{d}{dr} f^2 = 2r - 2 f^2 \left( \frac{r}{r^2 - \ell^2}
+ \frac{r}{r^2 + \ell^2}  + \frac{1}{r} \right)~.
\eeq
It is possible to integrate this differential equation;
\beq
f^2 = \frac{r^2}{4} \left( 1 - \frac{\ell^4}{r^4} \right) U(r)~,
\quad U(r) = 1 - \frac{k^8}{(r^4 - \ell^4)^2} ~.
\eeq
We thus find the following metric of $SU(4)$ holonomy;
\beqa
ds^2 &=& \left( 1 - \frac{\ell^4}{r^4} \right)^{-1} U(r)^{-1} dr^2
+ \frac{1}{2} (r^2 - \ell^2) (\sigma_1^2 + \sigma_2^2)
+ \frac{1}{2} (r^2 + \ell^2) (\Sigma_1^2 + \Sigma_2^2)  \CR
& &~~~~~+ r^2 (\tau_1^2 + \tau_2^2)
+ \frac{1}{4} r^2  \left( 1 - \frac{\ell^4}{r^4} \right) U(r) T_A^2~, \quad
( (k^4+\ell^4)^{1/4} \leq r)  \label{SU4}
\eeqa
which is asymptotically conical. The singular orbit at
$r=(k^4+\ell^4)^{1/4} \; (k \neq 0)$
is the flag manifold $SU(3)/T^2$, or the twister space of ${\bf CP}(2)$.
Hence this metric is a Ricci-flat K\"ahler metric on the canonical
line bundle over the flag manifold and it is in the class discussed
in \cite{BB}, \cite{PP}. (See also \cite{HKN} on the construction
of Ricci-flat metrics on the canonical line bundle over Hermitian
symmetric spaces based on the K\"ahler potential
of supersymmetric gauge theory.) We also note that
when $\ell=0, k \neq 0$, the metric constructed in \cite{CGLP2} is reproduced.
The first order system in \cite{CGLP2} corresponds to the case $a=b,
c_1=-c_2$ in
this paper and cannot cover the most general case. This is the reason why
we can obtain more general solutions with two parameters.
On the other hand, when $\ell \neq 0, k =0$ then $U(r) \equiv 1$ and
the solution reduces to the Calabi
hyperK\"ahler metric over $T^{*}{\bf CP}(2)$ of $Sp(2)$ holonomy \cite{DS}, 
\cite{CGLP2}~;
\beqa
ds^2 &=& \left( 1 - \frac{\ell^4}{r^4} \right)^{-1} dr^2
+ \frac{1}{2} (r^2 - \ell^2) (\sigma_1^2 + \sigma_2^2)
+ \frac{1}{2} (r^2 + \ell^2) (\Sigma_1^2 + \Sigma_2^2)  \CR
& &~~~~~+ r^2 (\tau_1^2 + \tau_2^2)
+ \frac{1}{4} r^2  \left( 1 - \frac{\ell^4}{r^4} \right) T_A^2~, \quad
(\ell \leq r ) \label{cara}
\eeqa
with three K\"ahler forms~,
\beqa
\omega^1 &=& \omega~, \CR
\omega^2 &=& cdt \wedge \tau_1 - cfT_{A} \wedge \tau_2 - ab(
\sigma_1 \wedge \Sigma_1 - \sigma_2 \wedge \Sigma_2)~, \\
\omega^3 &=& cdt \wedge \tau_2 + cfT_{A} \wedge \tau_1 + ab(
\sigma_1 \wedge \Sigma_2 + \sigma_2 \wedge \Sigma_1)~.
\nonumber
\eeqa
The flag manifold $SU(3)/T^2$ is a two-sphere bundle over ${\bf CP} (2)$ and
in the limit $k \to 0$ the $S^2$-fiber collapses to develop the singular orbit
${\bf CP} (2)$.

\section{Two types of ALC $Spin(7)$ metric}
\setcounter{equation}{0}

 From the view point of compactification of $M$ theory
it is of great interest to classify possible asymptotically
locally conical (ALC) metric. When $c_1 = c_2$,
an example of such ALC $Spin(7)$ metric is
given by \cite{KY}, \cite{CGLP5};
\beqa
ds^2 &=&  \frac {(r - \ell)^2}{(r + \ell)(r - 3\ell)} dr^2
+ (r - \ell)(r + \ell) (\sigma_1^2 + \sigma_2^2 +
\Sigma_1^2 + \Sigma_2^2) \CR
& &~~~~
+ (r - 3\ell)(r + \ell) (\tau_1^2 + \tau_2^2)
+ \ell^2\frac{(r + \ell)(r - 3\ell)}
{(r - \ell)^2}T_A^2~,  \quad (3\ell \leq r)
\eeqa
where the fiber over the base ${\bf CP}(2)$ is not ${\bf R}^4$ but
${\bf R}^4/{\bf Z}_2$. Note that this is different from the case of the Calabi
metric.
This is due to the fact that for the Calabi metric
it is $\sigma_1^2 + \sigma_2^2$ part that is collapsing at the singular
orbit, but it is $\tau_1^2 + \tau_2^2$ in the above ALC metric.

Let us consider the ALC $Spin(7)$ solutions that
interpolate between the short distance geometry of the form
${\bf R}^2 \times Flag_6$ (A-type boundary) or ${\bf R}^4 \times {\bf CP}(2)$
(B-type boundary) and the large distance geometry of the form
$S^1 \times C(Flag_6)$, where $C(Flag_6)$
is the 7-dimensional cone over $Flag_6$.
We then assume that the metric functions take the following form
for the large distance $t$~:
\begin{enumerate}
\item
$a(t)=ta_0+\alpha(t)$, $\;$
$b(t)=tb_0+\beta(t)$, $\;$
$c_1(t)=\gamma_1(t)$, \\
$c_2(t)=tc_{20}+\gamma_2(t)$, $\;$
$f(t)=tf_0+\zeta(t)$,
\item
$a(t)=ta_0+\alpha(t)$, $\;$
$b(t)=tb_0+\beta(t)$, \\
$c_1(t)=tc_{10}+\gamma_1(t)$, $\;$
$c_2(t)=tc_{20}+\gamma_2(t)$, $\;$
$f(t)=\zeta(t)$,
\end{enumerate}
where $a_0$, $b_0$, $c_{i0}$, $f_0$ are constants and
$\alpha$, $\beta$, $\gamma_i$, $\zeta$
are smooth functions tending to finite value for $t \rightarrow \infty$.
In case 1 the $S^1$ direction is $\tau_1$ and at large $t$ the function $c_1$
approaches a constant $M_1=\gamma_1(\infty)$, while in case 2 the
function $f$ of the $S^1$ direction $T_A$ approaches a constant
$M_2=\zeta(\infty)$. The octonionic instanton equation (\ref{five}) requires
the following conditions on the leading coefficients;
\beq
a_{0}^2=b_{0}^2=1, \;  a_0 b_0 c_{20}=1, \;  f_0=-1/2
\quad  \mbox{for case 1}~, \label{as1}
\eeq
\beq
a_{0}^2=b_{0}^2=1, \;  a_0 b_0 c_{20}=1, \;  c_{10}=c_{20}
\quad \mbox{for case 2}~. \label{as2}
\eeq
Note that they are different only in the last condition, but this 
difference produces
significant change as we will see shortly.
Thus the possible cone metrics on $C(Flag_6)$
consistent with the instanton equation are given by
\beqa
g_c &=& dt^2 + t^2(\sigma_1^2 + \sigma_2^2 + \Sigma_1^2 + \Sigma_2^2
+ \tau_2^2 + T_A^2/4)~, \\
g'_{c} &=& dt^2 + t^2(\sigma_1^2 + \sigma_2^2 + \Sigma_1^2 + \Sigma_2^2
+ \tau_1^2 + \tau_2^2)~,
\eeqa
corresponding to (\ref{as1}) and (\ref{as2}), respectively. It follows that
the boundary condition for the ALC metric is
\beq
g \rightarrow M_{1}^2 \tau_{1}^2+g_{c} \quad  \mbox{or} \quad
M_{2}^2 T_{A}^2+g'_{c}
\quad \mbox{for} \; t \rightarrow \infty.
\eeq
We call the first case $A_{\infty}^{\pm}$ and the second case $B_{\infty}$, and
prove the following proposition. The sign $\pm$ corresponds to the choice
$a_{0}=\pm b_{0}$ in (\ref{as1})\footnote{In (\ref{as2}) the choice of the sign
does not make difference due to the ${\bf Z}_2$ symmetry of the instanton 
equation.}~;
this difference does not appear at the
level of cone metrics, but it must be distinguished at the level of instanton
solutions.

\vskip0.8mm

\begin{flushleft}
{\bf Proposition} \quad If there exists a regular ALC solution interpolating
between $S^1 \times C(Flag_6)$ and ${\bf R}^2 \times Flag_6$ or
${\bf R}^4 \times {\bf CP}(2)$, then the following holds~:
\begin{enumerate}
\item
For the boundary $A_{\infty}^{\pm}$, $a(t)=\pm b(t)$ in the whole region
$0 \le t \le \infty$ and
the solution approaches ${\bf R}^2 \times Flag_6$ for $t \rightarrow 0$.
\item
For the boundary $B_{\infty}$, $c_1(t)=c_2(t)$ in the whole region
$0 \le t \le \infty$ and
the solution approaches
${\bf R}^4 \times {\bf CP}(2)$ for $t \rightarrow 0$.
\end{enumerate}
\end{flushleft}

\vskip0.8mm

\begin{flushleft}
{\bf Remark} \quad The part $dt^2+4t^2 T_A^2$ in the metric (\ref{ab})
looks like
\beq
dt^2+t^2 d\psi^2, \quad (0 \le \psi < 4\pi)
\eeq
when we fix the coordinates on $Flag_6$ in A-type boundary \cite{KY}.
Therefore, the range of $\psi$
must be adjusted to be that of usual polar coordinates on ${\bf R}^2$,
$0 \le \psi < 2\pi$. This means the manifold in the boundary
$A_{\infty}^{\pm}$ is $Flag_6/{\bf Z}_2$ rather than $Flag_6$, which
would have $0 \le \psi < 4\pi$. While in the case of $B_{\infty}$
it is not necessary to do such a modification since
\beq
dt^2+t^2(T_A^2+\sigma_1^2+\sigma_2^2)
\eeq
in (\ref{bb}) is the
standard metric on ${\bf R}^4$ written by the polar coordinates when we
fix the coordinates on ${\bf CP}(2)$.
\end{flushleft}

\vskip0.8mm

(Proof.) \quad We first consider the case $A_{\infty}^{+}$. From the
instanton equation
we have
\beqa
a(t) - b(t) &=& N \exp\left(\int^t u(t')dt' \right)~, \label{cc1}
\eeqa
where $N$ is an  integration constant and
\beq
u(t) = \frac{1}{2ab}\left(\frac{c_1^2-(a+b)^2}{c_1}+
\frac{c_2^2-(a+b)^2}{c_2} + 2f \right)~.
\eeq
Suppose that a regular solution exists
in the form (\ref{cc1}) with $N \neq 0$.
By using (\ref{as1}) with $a_0 = b_0$, it is easy to see the asymptotic
behavior
$a - b \simeq e^{-2t/M_{1}}$ for $t \rightarrow \infty$.
Note that the constant $M_1$ is required to be positive for
the exponentially small suppression.
If the solution approaches the singular
orbit $Flag_6$, then the product $c_1 c_2$ must be negative
by the result of section 2 (see also (\ref{sin1})). On the other hand,
$c_1 c_2$ is positive in the asymptotic
region since $c_1 c_2 \rightarrow M_1$ for $t \rightarrow \infty$
and hence $c_1$ or $c_2$ becomes zero at a certain time $t_0$,
which contradicts the regularity condition. If the solution approaches
the singular orbit ${\bf CP}(2)$, $f$ is positive as seen in
(\ref{sin2}), so the negative $f$ in the asymptotic region leads to
a contradiction. In the case of (\ref{sin3}), the product $c_1 c_2$
is negative for $t \rightarrow 0$,
which contradicts the sign in the asymptotic region.
Thus $a = b$ in the whole region, and if there exists a regular solution
with the boundary $A_{\infty}^{+}$, then it  approaches
${\bf R}^2 \times Flag_6$ for
$t \rightarrow 0$ since this boundary is only one consistent with $a=b$.
Furthermore, by the discrete symmetry of the instanton equation
there exists a regular solution with $a=-b$ for the boundary
$A_{\infty}^{-}$.

Next let us consider the case $B_{\infty}$. By the instanton equation
we have
\beq
c_{1}(t)-c_{2}(t)=N \exp\left(\int^t v(t')dt' \right)~,  \label{cc2}
\eeq
with
\beq
v(t)=\frac{4f^2-(c_1+c_2)^2}{2c_1 c_2 f}-\frac{c_1+c_2}{ab}.
\eeq
If we assume a regular solution (\ref{cc2}) with $N \neq 0$, then
similar arguments lead to a contradiction.
Thus $c_1=c_2$ in the whole region and the solution
approaches ${\bf R}^4 \times {\bf CP}(2)$ given by (\ref{sin2}) for
$t \rightarrow 0$.

\section{ Evidence for new global solutions}
\setcounter{equation}{0}

It is not easy to find exact solutions in general and we turn to numerical
computations to examine the existence of global solutions to
the octonionic instanton equation (\ref{five}).
The result of our analysis is summarized in Figure 1\footnote{The figure is attached
at the end of the section.}, which shows
possible lines for existence of global solutions in the two dimensional
parameter space of initial conditions at the singular orbit.
The circle $p^2+q^2=m^2$ represents the
Ricci-flat K\"ahler metrics of $SU(4)$ holonomy obtained in section 3.
As one can see from (\ref{SU4}), near the singular orbit at
$r_0=(k^4+\ell^4)^{1/4}$
we have
\beq
g \longrightarrow d\rho^2+4\rho^2 T_{A}^2 + \frac{1}{2}(r_0^2+\ell^2)
(\sigma_1^2+\sigma_2^2)
+\frac{1}{2}(r_0^2-\ell^2)(\Sigma_1^2+\Sigma_2^2)+
r_0^2(\tau_1^2+\tau_2^2)~,
\eeq
where $\rho^2=r_0(r-r_0)/2$. By comparing with the expansion (\ref{sin1}),
we see that they are
parametrized by the circle with radius $m=r_0$ in the $(p,q)$-space of
A-type boundary. The four points on the circle,
$(p,q)=(\pm m, 0)$ and $(0, \pm m)$,
correspond the Calabi hyperK\"ahler  metric given by (\ref{cara}),
where the holonomy group is
further reduced to $Sp(2)$. Note that the Calabi metric satisfies B-type
boundary and hence the singular orbit changes from $Flag_6$ to ${\bf CP}(2)$
at these points.
The wavy lines attaching to the Calabi metric are the ALC metrics
of $Spin(7)$ holonomy whose existence was expected by the second statement
of proposition ($B_{\infty}$ boundary). Indeed, from the numerical analysis
we can find non-singular solutions interpolating between $S^1 \times C(Flag_6)$
and ${\bf R}^4 \times {\bf CP}(2)$ for the parameter region $q_1 < -2/3$
of B-type boundary (\ref{sin2}), with $q_1=-2/3$ giving the Calabi metric
\cite{KY} \cite{CGLP5}.

Finally we discuss the new metrics of $Spin(7)$ holonomy depicted by the
lines $p=\pm q, p^2+q^2>m^2$ in Figure 1, which we shall denote by
${\bf C}_{8}^{*}$.
They are an analogue of $Spin(7)$
metrics ${\bf C}_8$ on the line bundle over ${\bf CP}(3)$
discussed in \cite{CGLP5} \cite{CGLP6}.
Although we have not been able to find the solutions in closed form,
the following arguments indicate they must exist.
The solutions on the two lines $p=q$ and $p=-q$ are related to each other
by the action of the discrete symmetry of the instanton equation
(\ref{five}), and so we will consider the case $p=q$. By rescaling the
parameter $p \rightarrow mp$, the perturtative expansion for
A-type boundary becomes
\beqa
a(t) &=& b(t)=m \left(p+\frac{6p^2-1}{4p^3}(t/m)^2+
\cdot \cdot \cdot \right)~, \CR
c_1(t) &=& m
\left(1+\frac{2p^2-1}{2p^2}(t/m)+\frac{12p^4-4p^2+3}{8p^4}(t/m)^2
+\cdot \cdot \cdot \right)~, \CR
c_2(t) &=& -m
\left(1-\frac{2p^2-1}{2p^2}(t/m)+\frac{12p^4-4p^2+3}{8p^4}(t/m)^2
+\cdot \cdot \cdot \right)~, \label{atype} \\
f(t) &=& -2t \left(1-\frac{12p^4+20p^2-1}{12p^4}(t/m)^2+
\cdot \cdot \cdot \right)~, \nonumber
\eeqa
which shows the reduction $a=b$ of the instanton equation.
If we put $c_3 \equiv -2 f$, the first order system with $a=b$ reduction 
is described by
\beqa
\frac{\dot a}{~a~} &=& \frac{c_1}{2a^2}
+ \frac{c_2}{2a^2} + \frac{c_3}{2a^2}~, \CR
\frac{\dot c_1}{~c_1~} &=& -\frac{c_1}{a^2}
+ \frac{c_1^2 - (c_2 -c_3)^2}{c_1 c_2 c_3}~, \label{ah} \\
\frac{\dot c_2}{~c_2~} &=& -\frac{c_2}{a^2}
+ \frac{c_2^2 - (c_3 -c_1)^2}{c_1 c_2 c_3}, \CR
\frac{\dot c_3}{~c_3~} &=& -\frac{c_3}{a^2}
+ \frac{c_3^2 - (c_1 -c_2)^2}{c_1 c_2 c_3}~.
\nonumber
\eeqa
After the rescaling $a \to \sqrt{2} a$ we obtain exactly the same first
order system as eq.(8) in \cite{CGLP5}, where
it was shown numerically that the solution with the boundary (\ref{atype})
is regular and ALC provided that the parameter $p$ is chosen to satisfy
$p^2 > 1/2$. It is easy to check the boundary $p^2=1/2$ corresponds to the 
AC solution
(\ref{SU4}) with the special value $\ell=0$. Thus two-parameter family of
ALC metrics
${\bf C}_{8}^{*}$ has the same topology as the canonical
line bundle over $Flag_6$. The large distance geometry of ${\bf C}_{8}^{*}$
can be worked out as follows. By the proposition in section 4,
${\bf C}_{8}^{*}$ approaches
the boundary $A_{\infty}^{+}$ for $t \rightarrow \infty$. After some
calculation we find that the asymptotic expansion up to order $t^{-3}$
is given by
\beqa
a(t) &=& t \left(1+\frac{3}{8}(M/t)^2+\frac{1}{4}(M/t)^3+
\frac{1}{2}\left(\frac{7}{64}-P \right)(M/t)^4+\cdot \cdot \cdot \right)~, \CR
c_1(t) &=& M\left(1-\frac{1}{2}(M/t)^2-\frac{1}{2}(M/t)^3+
\cdot \cdot \cdot \right)~, \label{series} \\
c_2(t) &=& c_3(t)=t\left(1-\frac{1}{2}(M/t)+
P(M/t)^4+ \cdot \cdot \cdot \right)~.
\nonumber
\eeqa
It should be noticed that the equality $c_2=c_3$ is valid for all orders,
if we assume that they can be expanded as power series in $t^{-1}$.
Hence, the series coincides with the asymptotic form of the ALC solutions
found in \cite{CGLP3}. The parameters $M,P$ correspond to $m,p$
in the expansion around the singular orbit.
Since the product $c_1 c_2=-m^2$ for $t \rightarrow 0$, we
must have $M < 0$. This sign is consistent with the exponentially small
correction of the asymptotic expansion. Indeed, we have
\beq
\dot c_2 - \dot c_3 =(c_2-c_3)\left(\frac{(c_2+c_3)^2-c_1^2}{c_1 c_2
c_3}-\frac{c_2+c_3}{a^2} \right)~,
\eeq
which leads to the asymptotic behavior $c_2-c_3 \simeq e^{4t/M}$ using the
expansion (\ref{series}),
and the metric functions behave similarly to those
in the Atiyah-Hitchin metric \cite{GM}, \cite{CGLP6}.


\newpage

\unitlength 0.1in
\begin{picture}( 33.9900, 35.7000)( 19.0100,-38.7000)
%
\special{pn 13}%
\special{ar 3600 2000 800 800  0.0000000 6.2831853}%
%
\special{pn 8}%
\special{pa 3600 1200}%
\special{pa 3566 1176}%
\special{pa 3550 1154}%
\special{pa 3568 1136}%
\special{pa 3606 1122}%
\special{pa 3642 1110}%
\special{pa 3648 1096}%
\special{pa 3618 1082}%
\special{pa 3576 1068}%
\special{pa 3552 1054}%
\special{pa 3564 1038}%
\special{pa 3604 1024}%
\special{pa 3640 1010}%
\special{pa 3648 996}%
\special{pa 3618 982}%
\special{pa 3578 968}%
\special{pa 3552 954}%
\special{pa 3564 938}%
\special{pa 3604 924}%
\special{pa 3640 910}%
\special{pa 3648 896}%
\special{pa 3618 882}%
\special{pa 3576 868}%
\special{pa 3552 852}%
\special{pa 3564 838}%
\special{pa 3604 824}%
\special{pa 3640 810}%
\special{pa 3648 796}%
\special{pa 3618 782}%
\special{pa 3576 768}%
\special{pa 3552 752}%
\special{pa 3566 738}%
\special{pa 3604 724}%
\special{pa 3642 710}%
\special{pa 3648 696}%
\special{pa 3618 682}%
\special{pa 3576 666}%
\special{pa 3552 652}%
\special{pa 3566 638}%
\special{pa 3604 624}%
\special{pa 3642 610}%
\special{pa 3648 596}%
\special{pa 3616 582}%
\special{pa 3576 566}%
\special{pa 3550 552}%
\special{pa 3566 538}%
\special{pa 3606 524}%
\special{pa 3642 510}%
\special{pa 3648 496}%
\special{pa 3616 480}%
\special{pa 3576 466}%
\special{pa 3550 452}%
\special{pa 3566 438}%
\special{pa 3604 424}%
\special{pa 3640 410}%
\special{pa 3648 396}%
\special{pa 3622 380}%
\special{pa 3582 366}%
\special{pa 3552 352}%
\special{pa 3552 338}%
\special{pa 3578 324}%
\special{pa 3620 310}%
\special{pa 3650 300}%
\special{sp}%
%
\special{pn 8}%
\special{pa 3600 2800}%
\special{pa 3636 2824}%
\special{pa 3652 2846}%
\special{pa 3634 2864}%
\special{pa 3594 2878}%
\special{pa 3560 2892}%
\special{pa 3554 2904}%
\special{pa 3582 2918}%
\special{pa 3624 2932}%
\special{pa 3650 2948}%
\special{pa 3636 2962}%
\special{pa 3598 2976}%
\special{pa 3560 2990}%
\special{pa 3552 3004}%
\special{pa 3582 3020}%
\special{pa 3624 3034}%
\special{pa 3650 3048}%
\special{pa 3636 3062}%
\special{pa 3598 3076}%
\special{pa 3560 3090}%
\special{pa 3552 3106}%
\special{pa 3582 3120}%
\special{pa 3624 3134}%
\special{pa 3650 3148}%
\special{pa 3636 3162}%
\special{pa 3598 3176}%
\special{pa 3560 3190}%
\special{pa 3554 3206}%
\special{pa 3584 3220}%
\special{pa 3624 3234}%
\special{pa 3650 3248}%
\special{pa 3636 3262}%
\special{pa 3596 3276}%
\special{pa 3560 3292}%
\special{pa 3554 3306}%
\special{pa 3584 3320}%
\special{pa 3626 3334}%
\special{pa 3650 3348}%
\special{pa 3636 3362}%
\special{pa 3596 3376}%
\special{pa 3560 3392}%
\special{pa 3554 3406}%
\special{pa 3584 3420}%
\special{pa 3626 3434}%
\special{pa 3650 3448}%
\special{pa 3636 3462}%
\special{pa 3596 3478}%
\special{pa 3560 3492}%
\special{pa 3554 3506}%
\special{pa 3584 3520}%
\special{pa 3626 3534}%
\special{pa 3650 3548}%
\special{pa 3636 3564}%
\special{pa 3596 3578}%
\special{pa 3560 3592}%
\special{pa 3552 3606}%
\special{pa 3578 3620}%
\special{pa 3618 3634}%
\special{pa 3648 3648}%
\special{pa 3650 3664}%
\special{pa 3622 3678}%
\special{pa 3580 3692}%
\special{pa 3550 3700}%
\special{sp}%
%
\special{pn 8}%
\special{pa 4400 2000}%
\special{pa 4424 1966}%
\special{pa 4446 1948}%
\special{pa 4464 1966}%
\special{pa 4478 2004}%
\special{pa 4492 2040}%
\special{pa 4504 2048}%
\special{pa 4518 2018}%
\special{pa 4532 1976}%
\special{pa 4548 1950}%
\special{pa 4562 1962}%
\special{pa 4576 2002}%
\special{pa 4590 2038}%
\special{pa 4604 2048}%
\special{pa 4618 2018}%
\special{pa 4634 1978}%
\special{pa 4648 1950}%
\special{pa 4662 1962}%
\special{pa 4676 2002}%
\special{pa 4690 2040}%
\special{pa 4704 2048}%
\special{pa 4720 2018}%
\special{pa 4734 1976}%
\special{pa 4748 1950}%
\special{pa 4762 1964}%
\special{pa 4776 2002}%
\special{pa 4790 2040}%
\special{pa 4806 2046}%
\special{pa 4820 2016}%
\special{pa 4834 1974}%
\special{pa 4848 1950}%
\special{pa 4862 1964}%
\special{pa 4876 2002}%
\special{pa 4892 2040}%
\special{pa 4906 2046}%
\special{pa 4920 2016}%
\special{pa 4934 1974}%
\special{pa 4948 1948}%
\special{pa 4962 1964}%
\special{pa 4978 2004}%
\special{pa 4992 2040}%
\special{pa 5006 2046}%
\special{pa 5020 2014}%
\special{pa 5034 1974}%
\special{pa 5048 1948}%
\special{pa 5062 1964}%
\special{pa 5078 2004}%
\special{pa 5092 2040}%
\special{pa 5106 2046}%
\special{pa 5120 2014}%
\special{pa 5134 1974}%
\special{pa 5148 1948}%
\special{pa 5162 1962}%
\special{pa 5176 2000}%
\special{pa 5192 2038}%
\special{pa 5206 2046}%
\special{pa 5220 2020}%
\special{pa 5234 1980}%
\special{pa 5248 1950}%
\special{pa 5262 1948}%
\special{pa 5278 1974}%
\special{pa 5292 2016}%
\special{pa 5300 2048}%
\special{sp}%
%
\special{pn 8}%
\special{pa 2800 2000}%
\special{pa 2776 2036}%
\special{pa 2754 2052}%
\special{pa 2736 2034}%
\special{pa 2722 1994}%
\special{pa 2710 1958}%
\special{pa 2698 1952}%
\special{pa 2684 1982}%
\special{pa 2668 2022}%
\special{pa 2654 2048}%
\special{pa 2640 2036}%
\special{pa 2626 1998}%
\special{pa 2612 1960}%
\special{pa 2598 1952}%
\special{pa 2582 1980}%
\special{pa 2568 2022}%
\special{pa 2554 2048}%
\special{pa 2540 2036}%
\special{pa 2526 1998}%
\special{pa 2512 1960}%
\special{pa 2498 1950}%
\special{pa 2484 1978}%
\special{pa 2468 2020}%
\special{pa 2454 2048}%
\special{pa 2440 2036}%
\special{pa 2426 1998}%
\special{pa 2412 1960}%
\special{pa 2398 1950}%
\special{pa 2382 1980}%
\special{pa 2368 2020}%
\special{pa 2354 2048}%
\special{pa 2340 2036}%
\special{pa 2326 1996}%
\special{pa 2312 1958}%
\special{pa 2298 1950}%
\special{pa 2284 1978}%
\special{pa 2268 2020}%
\special{pa 2254 2046}%
\special{pa 2240 2036}%
\special{pa 2226 1996}%
\special{pa 2212 1958}%
\special{pa 2198 1950}%
\special{pa 2182 1978}%
\special{pa 2168 2020}%
\special{pa 2154 2046}%
\special{pa 2140 2034}%
\special{pa 2126 1996}%
\special{pa 2112 1958}%
\special{pa 2098 1948}%
\special{pa 2084 1976}%
\special{pa 2068 2018}%
\special{pa 2054 2046}%
\special{pa 2040 2036}%
\special{pa 2026 1998}%
\special{pa 2012 1960}%
\special{pa 1998 1948}%
\special{pa 1982 1970}%
\special{pa 1968 2010}%
\special{pa 1954 2042}%
\special{pa 1940 2046}%
\special{pa 1926 2022}%
\special{pa 1912 1982}%
\special{pa 1902 1946}%
\special{sp}%
%
\special{pn 8}%
\special{pa 3030 1430}%
\special{pa 2030 430}%
\special{fp}%
\special{pa 4170 1430}%
\special{pa 5170 430}%
\special{fp}%
%
\special{pn 8}%
\special{pa 3030 2560}%
\special{pa 2030 3560}%
\special{fp}%
\special{pa 4170 2560}%
\special{pa 5170 3560}%
\special{fp}%
%
\special{pn 13}%
\special{sh 1}%
\special{ar 4400 1990 10 10 0  6.28318530717959E+0000}%
\special{sh 1}%
\special{ar 3576 1202 10 10 0  6.28318530717959E+0000}%
\special{sh 1}%
\special{ar 2788 2026 10 10 0  6.28318530717959E+0000}%
\special{sh 1}%
\special{ar 3612 2814 10 10 0  6.28318530717959E+0000}%
\special{sh 1}%
\special{ar 3612 2814 10 10 0  6.28318530717959E+0000}%
%
\special{pn 20}%
\special{pa 4130 2520}%
\special{pa 4210 2600}%
\special{fp}%
\special{pa 4210 2520}%
\special{pa 4130 2600}%
\special{fp}%
%
\special{pn 20}%
\special{pa 4130 1390}%
\special{pa 4210 1470}%
\special{fp}%
\special{pa 4210 1390}%
\special{pa 4130 1470}%
\special{fp}%
%
\special{pn 20}%
\special{pa 2990 1390}%
\special{pa 3070 1470}%
\special{fp}%
\special{pa 3070 1390}%
\special{pa 2990 1470}%
\special{fp}%
%
\special{pn 20}%
\special{pa 2990 2510}%
\special{pa 3070 2590}%
\special{fp}%
\special{pa 3070 2510}%
\special{pa 2990 2590}%
\special{fp}%
%
\special{pn 8}%
\special{pa 3100 2000}%
\special{pa 4100 2000}%
\special{fp}%
\special{sh 1}%
\special{pa 4100 2000}%
\special{pa 4034 1980}%
\special{pa 4048 2000}%
\special{pa 4034 2020}%
\special{pa 4100 2000}%
\special{fp}%
\special{pa 3600 2500}%
\special{pa 3600 1500}%
\special{fp}%
\special{sh 1}%
\special{pa 3600 1500}%
\special{pa 3580 1568}%
\special{pa 3600 1554}%
\special{pa 3620 1568}%
\special{pa 3600 1500}%
\special{fp}%
\put(41.0000,-21.0000){\makebox(0,0){$p$}}%
\put(37.0000,-15.0000){\makebox(0,0){$q$}}%
\put(50.0000,-10.0000){\makebox(0,0){$Spin(7)$}}%
\put(32.0000,-8.0000){\makebox(0,0){$Spin(7)$}}%
\put(43.5000,-22.3000){\makebox(0,0)[lt]{$SU(4)$}}%
\put(39.0000,-9.0000){\makebox(0,0){$Sp(2)$}}%
%
\special{pn 4}%
\special{pa 3850 1050}%
\special{pa 3610 1180}%
\special{fp}%
\special{sh 1}%
\special{pa 3610 1180}%
\special{pa 3678 1166}%
\special{pa 3658 1156}%
\special{pa 3660 1132}%
\special{pa 3610 1180}%
\special{fp}%
\put(34.1000,-21.7000){\makebox(0,0)[lb]{$0$}}%
\put(10.6000,-40.4000){\makebox(0,0)[lb]{Figure 1: 
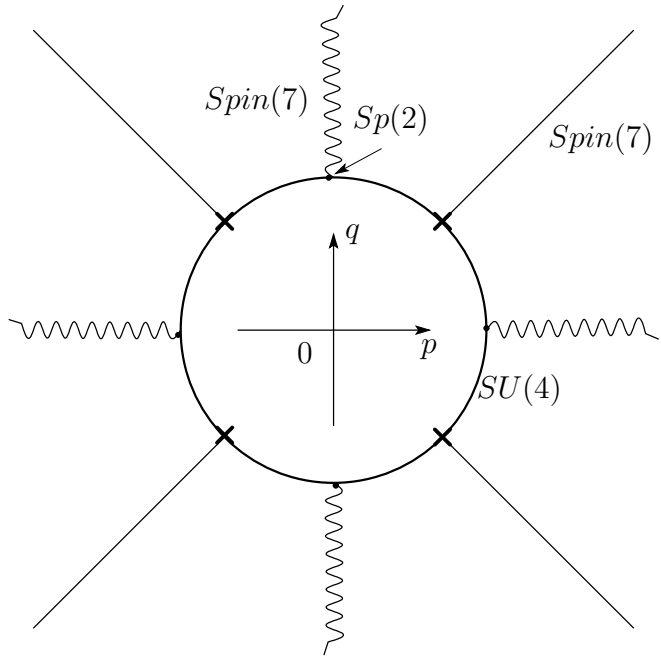
Possible lines for the existence of global metric of special holonomy.}}%
\end{picture}%

\newpage


\vskip10mm

\begin{center}
{\bf Acknowledgements}
\end{center}

We would like to thank G.W. Gibbons and C.N. Pope for correspondence.
This work is supported in part by the fund
for special priority area 707
"Supersymmetry and Unified Theory of Elementary Particles" and
the Grant-in-Aid for Scientific Research No. 12640074.



\section*{Appendix A}
\renewcommand{\theequation}{A.\arabic{equation}}\setcounter{equation}{0}
\begin{flushleft}
{\bf Convention of $SU(3)$ Maurer-Cartan forms}
\end{flushleft}

We use the following
$SU(3)$ Maurer-Cartan equation that is $\Sigma_3$ symmetric;
\beqa
d\sigma_1 &=& \Sigma_1 \wedge \tau_1 - \Sigma_2 \wedge \tau_2 + \kappa_A
T_A \wedge \sigma_2
+ \kappa_B T_B \wedge \sigma_2~, \CR
d\sigma_2 &=& - \Sigma_1 \wedge \tau_2 - \Sigma_2 \wedge \tau_1 - \kappa_A
T_A \wedge \sigma_1
- \kappa_B T_B \wedge \sigma_1~, \CR
d\Sigma_1 &=& \tau_1 \wedge \sigma_1 - \tau_2 \wedge \sigma_2 + \mu_A T_A
\wedge \Sigma_2
+ \mu_B T_B \wedge \Sigma_2~, \CR
d\Sigma_2 &=& - \tau_1 \wedge \sigma_2 - \tau_2 \wedge \sigma_1 - \mu_A T_A
\wedge \Sigma_1
- \mu_B T_B \wedge \Sigma_1~, \\
d\tau_1 &=& \sigma_1 \wedge \Sigma_1 - \sigma_2 \wedge \Sigma_2 + \nu_A T_A
\wedge \tau_2
+ \nu_B T_B \wedge \tau_2~, \CR
d\tau_2 &=& - \sigma_1 \wedge \Sigma_2 - \sigma_2 \wedge \Sigma_1 - \nu_A
T_A \wedge \tau_1
- \nu_B T_B \wedge \tau_1~, \CR
dT_A &=& 2\alpha_A \sigma_1 \wedge \sigma_2  + 2\beta_A \Sigma_1 \wedge
\Sigma_2
+ 2\gamma_A \tau_1 \wedge \tau_2~, \CR
dT_B &=& 2\alpha_B \sigma_1 \wedge \sigma_2  + 2\beta_B \Sigma_1 \wedge
\Sigma_2
+ 2\gamma_B \tau_1 \wedge \tau_2~. \nonumber
\eeqa
This form of the Maurer-Cartan equation is symmetric under the (cyclic)
permutation of $(\sigma_i, \Sigma_i,
\tau_i)$.
 From the Jacobi identity we see that
the parameters
$\alpha, \beta, \gamma, \kappa, \mu,\nu$, which describe the "coupling" of
the Cartan generators $\{ T_A, T_B\}$ satisfy
\beqa
& & \alpha_A + \beta_A + \gamma_A = 0~, \;
\alpha_B + \beta_B + \gamma_B = 0~,  \CR
\kappa_A &=& \frac{1}{\Delta}(\beta_B - \gamma_B), \;
\kappa_B = -\frac{1}{\Delta}(\beta_A - \gamma_A), \;
\mu_A = -\frac{1}{\Delta}(\alpha_B-\gamma_B), \\
\mu_B &=& \frac{1}{\Delta}(\alpha_A - \gamma_A), \;
\nu_A = \frac{1}{\Delta}(\alpha_B - \beta_B), \;
\nu_B = -\frac{1}{\Delta}(\alpha_A - \beta_A) \nonumber
\eeqa
with $\Delta=\beta_A \alpha_B -\alpha_A \beta_B$
leaving four free parameters $(\alpha_{A,B}, \beta_{A,B})$.
We may further put the "orthogonality" conditions;
\beqa
\alpha_A \alpha_B + \beta_A \beta_B + \gamma_A \gamma_B &=& 0~, \CR
\kappa_A \kappa_B + \mu_A \mu_B + \nu_A \nu_B &=& 0~,
\eeqa
which reduces one parameter.

\newpage

\section*{Appendix B}
\renewcommand{\theequation}{B.\arabic{equation}}\setcounter{equation}{0}
\begin{flushleft}
{\bf Reduction of holonomy group and the octonionic instanton equation}
\end{flushleft}

One of ways to realize the reduction of holonomy group is to impose
appropriate linear relations on the $so(n)$ valued spin connection one form
$\omega_{ab} = - \omega_{ba}$. It is rather amusing that in all the three cases
which are most relevant from the viewpoint of $M$ theory conpactifications
the expected number of linear relations is always seven, since
${\rm dim}~SO(8) - {\dim}~Spin(7) = {\rm dim}~SO(7) - {\dim}~G_2=
{\rm dim}~SO(6) - {\dim}~SU(3) = 7$.  There are topological relations
behind this dimension counting; $Spin(7)/G_2 \simeq SO(8)/SO(7) \simeq S^7$ and
$G_2/SU(3) \simeq SO(7)/SO(6) \simeq S^6$.
In fact the following octonionic instanton equation gives
a \lq\lq master\rq\rq\ equation for seven conditions required for the 
reduction \cite{CDFN}, \cite{AL}.
\beq
\omega_{ab} = \frac{1}{2} \Psi_{abcd} \omega_{cd}~, \label{octduality}
\eeq
where totally anti-symmetric tensor $\Psi_{abcd}$ is defined by
the structure constants of octonions $\psi_{abc}$ as follows;
\beqa
\Psi_{abc0} &=& \psi_{abc}~,  \quad (1 \leq a,b,c, \cdots \leq 7) \CR
\Psi_{abcd} &=& -\frac{1}{3!} \epsilon_{abcdefg} \psi_{efg}~.
\eeqa
A conventional choice of the structure constants is
\beqa
\psi_{abc}= +1~, ~~~{\rm for}~~~(abc)=(123), (516), (624), (435), (471),
(572), (673)~.
\eeqa
It can be shown that (\ref{octduality}) implies the four form defined by
\beq
\Omega = \frac{1}{4!}
\Psi_{abcd} e^a \wedge e^b \wedge e^c \wedge
e^d~. \label {calib}
\eeq
is closed and the metric has $Spin(7)$ holonomy \cite{BFK}.
In the above convention of the structure constants of octonions
the explicit form of the octonionic instaton equation is
\beqa
\omega_{14} + \omega_{25} + \omega_{36}  + \omega_{07} &=& 0~, \CR
\omega_{71} + \omega_{62} + \omega_{35}  + \omega_{04} &=& 0~, \CR
\omega_{47} + \omega_{65} + \omega_{23}  + \omega_{01} &=& 0~, \CR
\omega_{67} + \omega_{12} + \omega_{54}  + \omega_{03} &=& 0~, 
\label{explicit} \\
\omega_{73} + \omega_{51} + \omega_{24}  + \omega_{06} &=& 0~, \CR
\omega_{57} + \omega_{46} + \omega_{31}  + \omega_{02} &=& 0~, \CR
\omega_{72} + \omega_{16} + \omega_{43}  + \omega_{05} &=& 0~. \nonumber
\eeqa
If we simply substitute $\omega_{0k},(1\leq k \leq 7)$, then 
(\ref{explicit}) gives
the seven conditions for $G_2$ holonomy.
Further putting $\omega_{7j},(1\leq j \leq 6)$ gives the seven conditions for
$SU(3)$ holonomy. We should emphasize that compared with the condition on
the Riemann curvature, the condition on the spin connection depends on the gauge
or the choice of coordinate system and therefore it is only a sufficient 
but not necessary condition for special holonomy.

\section*{Appendix C}
\renewcommand{\theequation} {C.\arabic{equation}}\setcounter {equation} {0}
\begin{flushleft}
{\bf Perturbative expansion around singular orbits}
\end{flushleft}

For A-type boundary the instanton equation is perturbatively solved
in the form,
\beqa
a(t) &=& p +\left(\frac{1}{p}+\frac{(p^2+q^2-m^2)(p^2-q^2+m^2)}
{4 p q^2 m^2} \right)t^2+ \cdot \cdot \cdot , \CR
b(t) &=& q +\left(\frac{1}{q}+\frac{(p^2+q^2-m^2)(-p^2+q^2+m^2)}
{4 p^2 q m^2} \right)t^2+ \cdot \cdot \cdot , \CR
c_{1}(t) &=& m+ \left(\frac{p^2+q^2-m^2}{2pq} \right)t
+\left(\frac{2}{m}-\frac{m(p^2+q^2-m^2)}{2p^2 q^2}
-\frac{(p^2+q^2-m^2)^2}{8p^2 q^2 m} \right)t^2+ \cdot \cdot \cdot ,
\label{sin1} \CR
c_{2}(t) &=& -m + \left(\frac{p^2+q^2-m^2}{2pq} \right)t
-\left(\frac{2}{m}-\frac{m(p^2+q^2-m^2)}{2p^2 q^2}
-\frac{(p^2+q^2-m^2)^2}{8p^2 q^2 m} \right)t^2+ \cdot \cdot \cdot , \CR
f(t) &=& -2t\left(1+
\left(\frac{p^4+q^4+m^4-10 p^2 m^2-10 q^2 m^2-14 p^2 q^2}{12 p^2 q^2 m^2}
\right)t^2+ \cdot \cdot \cdot  \right)~.
\eeqa
We note that the power series solution are completely fixed by
the "initial conditions" $p,q,m$. (This should be compared with
the case of B type boundary condition in the following.)
The reduction $c_1=-c_2$ is
reproduced by imposing $p^2 + q^2 = m^2$
and the reduction $a=\pm b$ by $p=\pm q$.

There are two possible solutions for B-type boundary. One of these is given by
\beqa
a(t) &=& t \left(1-\frac{1}{2}(q_{1}+1)(t/m)^2+ \cdot \cdot \cdot \right),
\CR
b(t) &=& m \left(1+\frac{1}{2}(t/m)^2+ \cdot \cdot \cdot \right),
\CR
c_{1}(t) &=& c_{2}(t)=m\left(1+(t/m)^2+ \cdot \cdot \cdot \right),
\label{sin2} \\
f(t) &=& t \left(1+q_{1}(t/m)^2+ \cdot \cdot \cdot \right).
\nonumber
\eeqa
The other solution has the following expansion~,
\beqa
a(t) &=& t \left(1-\frac{1}{6}(t/m)^2+ \cdot \cdot \cdot \right), \CR
b(t) &=& m \left(1+q_{2}(t/m)^2+ \cdot \cdot \cdot \right), \CR
c_{1}(t) &=& m \left(1+(t/m)^2+ \cdot \cdot \cdot \right), \label{sin3} \\
c_{2}(t) &=& -m\left(1+2(1-q_{2})(t/m)^2+ \cdot \cdot \cdot \right), \CR
f(t) &=& t \left(-1+ \frac{2}{3} (t/m)^2 + \cdot \cdot \cdot  \right).
\nonumber
\eeqa
These solutions include the free parameters $q_1$ and $q_2$ in addition to
the scaling parameter $m$. In particular,
both solutions with $q_{1}=-2/3$ and $q_{2}=1/2$ lead to a same metric
and this is in fact precisely the Calabi hyperK\"ahler metric on
$T^{*}{\bf CP}(2)$.




\end{document}